\documentclass[12pt]{article}
\usepackage{graphicx}
\usepackage[cp1251]{inputenc}
\usepackage[russian]{babel}
\usepackage{rotating}
\begin{document}
 \noindent {\footnotesize\it Astronomy Letters, 2017, Vol. 43, No 5, pp. 304--315.}
 \newcommand{\dif}{\textrm{d}}

 \noindent
 \begin{tabular}{llllllllllllllllllllllllllllllllllllllllllllll}
 & & & & & & & & & & & & & & & & & & & & & & & & & & & & & & & & & & & & & \\\hline\hline
 \end{tabular}

  \vskip 0.5cm
 \centerline{\bf\large Vertical Distribution and Kinematics of Planetary Nebulae}
 \centerline{\bf\large in the Milky Way}
 \bigskip
 \bigskip
  \centerline
 {
 V.V. Bobylev\footnote [1]{e-mail: vbobylev@gao.spb.ru} and
 A.T. Bajkova
 }
 \bigskip
 {
 \small\it
 Central (Pulkovo) Astronomical Observatory, Russian Academy of Sciences,

 Pulkovskoe sh. 65, St. Petersburg, 196140 Russia
 }
 \bigskip
 \bigskip
 \bigskip

 {
{\bf Abstract}---Based on published data, we have produced a
sample of planetary nebulae (PNe) that is complete within 2~kpc of
the Sun. We have estimated the total number of PNe in the Galaxy
from this sample to be $17000\pm3000$ and determined the vertical
scale height of the thin disk based on an exponential density
distribution to be $197\pm10$~pc. The next sample includes PNe
from the Stanghellini--Haywood catalog with minor additions. For
this purpose, we have used $\sim$200 PNe with Peimbert's types I,
II, and III. In this case, we have obtained a considerably higher
value of the vertical scale height that increases noticeably with
sample radius. We have experimentally found that it is necessary
to reduce the distance scale of this catalog approximately by
20\%. Then, for example, for PNe with heliocentric distances less
than 4 kpc the vertical scale height is $256\pm12$~kpc. A
kinematic analysis has confirmed the necessity of such a reduction
of the distance scale.
  }

 \subsection*{INTRODUCTION}
Planetary nebulae (PNe) reflect a very short ($10^3-10^4$~yr)
phase in the evolution of stars with a mass of $1-9 M_\odot.$ This
phase begins when an asymptotic giant branch (AGB) star ejects its
envelope and ends with the formation of a white dwarf. PNe in the
Galaxy are represented everywhere, in the thin and thick disks, in
the halo and the bulge, of course, in different proportions.
Therefore, PNe are an important source of information about the
structure of the Galaxy, its chemical and dynamical evolution.

More than 1500 PNe have been discovered in the Galaxy to date.
However, the estimates of their total number $N_{tot}$ differ
significantly, from $4\times10^3$ to $4\times10^5:$ 4000--22 000
(Alloin et al. 1976), $10000\pm4000$ (Jacoby 1980), 40000 (Amnuel
et al. 1984), $>10^5$ (Ishida and Weinberger 1987),
$(2-4)\times10^5$ (Khromov 1989), $7200\pm1800$ (Peimbert 1990),
$23000\pm6000$ (Zijlstra and Pottasch 1991), $46 000\pm13 000$
(Moe and Marco 2006), or $24000\pm4000$ (Frew 2008).

At present, there are measurements of the trigonometric parallaxes
for the central stars of PNe. Such measurements at optical
wavelengths have been performed on the basis of observations from
the ground (Harris et al. 2007) and spacecraft, in particular, the
Hubble Space Telescope (Benedict et al. 2009) and the Gaia
satellite (Gaia Collaboration, Prusti et al. 2016). The
trigonometric parallaxes of several PNe have also been measured
with ground-based very-long-baseline interferometers, in
particular, the parallax of the PN~K~3--35 was measured within the
Japanese VERA (VLBI Exploration of Radio Astrometry) Program
(Tafoya et al. 2011). As Stanghellini et al. (2016) showed, in the
first published Gaia Data Release (DR1, Gaia Collaboration, Brown
et al. 2016) the parallaxes of only seven PNe have been measured
so far with a relative error $<80\%.$ Of course, upon completion
of the satellite flight the number of reliably measured parallaxes
for PNe will increase several-fold.

Since highly accurate (with an error of 10--15\%) parallax
measurements cover very few PNe, the distances to these objects
are usually estimated by various indirect methods. In particular,
the statistical method by Shklovskii (1956), which is based on an
empirical relation between the ionized mass in the shell and its
radius, is widely used. In this case, the errors in the distances
are fairly large. An overview of the methods and a comparison of
various PN distance scales can be found in Smith (2015).

The Galactic disk is a complex formation that contains stars of
various ages belonging to various dynamical structures. It is
customary to divide it into the thin and thick disks by either
spatial or kinematic criteria. Several models of the Galaxy with
fixed values of, for example, the densities, the radial scale
lengths or scale heights for each disk have been constructed with
such an approach (Robin et al. 2003; Girardi et al. 2005). A more
complex approach suggesting that the disk consists of a multitude
of independently evolving structures, each having elemental
abundances in a very narrow range of values, has been developed in
recent years (Rix and Bovy 2013; Bovy et al. 2016). From this
viewpoint, analyzing the spatial distribution of such highly
specialized objects as PNe is of great interest.

The goal of this paper is to study the $z$ distribution and
kinematics of PNe in the Galaxy. This suggests solving the
following problems: producing a complete sample, estimating the
vertical disk scale height, estimating the total number of PNe in
the Galaxy, and studying the vertical distribution and kinematics
of PNe based on a sample from a wide solar neighborhood.

 \subsection*{METHODS}
 \subsubsection*{Vertical Distribution of Galactic Objects}
We use the heliocentric rectangular $xyz$ coordinate system where
the $x$ axis is directed toward the Galactic center, the $y$ axis
is in the direction of Galactic rotation, and the $z$ axis is
directed to the north Galactic pole. The heliocentric distance is
 $r^2 = x^2 + y^2 + z^2$ and the cylindrical radius is
 $d^2_{xy}=x^2+y^2.$

The observed frequency distribution of objects along the $z$
coordinate axis is described by expressions of the following from
in the model of an exponential density distribution:
 \begin{equation}
  N(z)=N_1 \exp \biggl(-{|z-z_\odot|\over h_1}\biggr),
 \label{exponent}
 \end{equation}
where $N_1$ is the normalization coefficient, $z_\odot$ is the
distance from the Sun to the Galactic midplane (the mean of the
$z$ coordinates of all objects from the sample), and $h_1$ is the
vertical disk scale height; in the model of a self-gravitating
isothermal disk (Conti and Vacca 1990)
 \begin{equation}
  N(z)=N_2{\hbox { sech}}^2 \biggl({z-z_\odot\over \sqrt2~h_2}\biggr),
 \label{self-grav}
 \end{equation}
where $N_2$ is the normalization coefficient, $h_2$ is the
vertical disk scale height; and, finally, in the Gaussian model
 \begin{equation}
  N(z)=N_3\exp\biggl[-{1\over 2}\biggl({z-z_\odot\over h_3}\biggr)^2\biggr],
 \label{Gauss}
 \end{equation}
where $N_3$ is the normalization coefficient. Although the
standard of the normal distribution $h_3$ in the strict sense
(this term is usually applicable to exponential laws) is not a
parameter of the vertical scale height, here we attach this
meaning to it.

To estimate the total number of objects in the Galaxy $N_{tot},$
we use the following relation (Piskunov et al. 2006):
 \begin{equation}
 N_{tot}=2\pi\Sigma_{xy}
 \int^{R^{lim}}_0 \exp\biggl(-{R-R_\odot\over L_d}\biggr)R dR,
 \label{Total-NNZ}
 \end{equation}
where $R$ is the projection of the distance from the star to the
Galactic center onto the Galactic $xy$ plane; $R^{lim}$ is the
Galactocentric radius of the disk; $L_d$ is the radial disk scale
length; $\Sigma_{xy}=N_f/(\pi d^2_{xy})$ is the surface density in
the solar neighborhood found under the condition of sample
completeness, where $d_{xy}$ is the sample completeness radius in
the $xy$ plane and $N_f$ is the number of objects in the sample.
We take $R^{lim}=15$~kpc, $L_d=3.5$~kpc, and $R_\odot=8$~kpc,
which correspond to the model from Bahcall and Soneira (1980).

 \subsubsection*{Kinematics}
To determine the parameters of the Galactic rotation curve, we use
the equations derived from Bottlinger’s formulas in which the
angular velocity $\Omega$ is expanded into a series to terms of
the second order of smallness in $r/R_0:$
\begin{equation}
 \begin{array}{lll}
 V_r=-U_\odot\cos b\cos l-V_\odot\cos b\sin l-W_\odot\sin b+\\
 +R_0(R-R_0)\sin l\cos b\Omega^{'}_0+0.5R_0(R-R_0)^2\sin l\cos b\Omega^{''}_0,
 \label{EQ-1}
 \end{array}
 \end{equation}
 \begin{equation}
 \begin{array}{lll}
 V_l= U_\odot\sin l-V_\odot\cos l-r\Omega_0\cos b+\\
 +(R-R_0)(R_0\cos l-r\cos b)\Omega^{'}_0+0.5(R-R_0)^2(R_0\cos l-r\cos b)\Omega^{''}_0,
 \label{EQ-2}
 \end{array}
 \end{equation}
 \begin{equation}
 \begin{array}{lll}
 V_b=U_\odot\cos l\sin b + V_\odot\sin l \sin b-W_\odot\cos b-\\
 -R_0(R-R_0)\sin l\sin b\Omega^{'}_0-0.5R_0(R-R_0)^2\sin l\sin b\Omega^{''}_0,
 \label{EQ-3}
 \end{array}
 \end{equation}
where $U_\odot,V_\odot,W_\odot$ are the components of the group
velocity vector for the stars being considered relative to the
local standard of rest (taken with the opposite sign), $R$ is the
distance from the star to the Galactic rotation axis:
  \begin{equation}
 R^2=r^2\cos^2 b-2R_0 r\cos b\cos l+R^2_0.
  \end{equation}
$\Omega_0$ is the angular velocity of Galactic rotation at the
Galactocentric distance $R_0$ of the Sun, the parameters
$\Omega^{'}_0$ and $\Omega^{''}$ are the corresponding derivatives
of the angular velocity, $V_0=|R_0\Omega_0|$. The velocities $V_R$
and $\Delta V_{circ}$ must be freed from the peculiar solar
velocity $U_\odot,V_\odot,W_\odot$.

It is important to know the specific value of the distance R0.
Gillessen et al. (2009) obtained one of its most reliable
estimates, $R_0=8.28\pm0.29$~kpc, by analyzing the orbits of stars
moving around the massive black hole at the Galactic center. The
following estimates were obtained from various samples of masers
with measured trigonometric parallaxes:
 $R_0=8.34\pm0.16$~kpc (Reid et al. 2014),
 $8.3\pm0.2$~kpc (Bobylev and Bajkova 2014),
 $8.03\pm0.12$~kpc (Bajkova and Bobylev 2015), and
 $8.40\pm0.12$~kpc (Rastorguev et al. 2016). Based on these results, we adopted
 $R_0=8.3\pm0.2$~kpc in this paper.

Note that we apply the kinematic equations (5)--(7) to the PNe
that are involved in the Galactic rotation. Therefore, we seek to
maximally free the sample from the noise introduced, for example,
by the halo objects that are no involved in this rotation.

 \subsection*{DATA}
The catalog by Stanghellini and Haywood (2010), which contains 728
PNe with distance estimates, served as one of the data sources for
our study. It is one of the largest catalogs of PNe with
homogeneous distance estimates to date. For example, the
bibliographic compilation by Acker et al. (1992) contains only 296
PNe with more or less reliable distance estimates obtained by
various methods. The catalog by Phillips (2004) numbers already
447 such nebulae. A historical background on the catalogs of PNe
can be found, for example, in Parker et al. (2006). PNe from the
catalog by Stanghellini and Haywood (2010) cover much of the
Galaxy; the heliocentric distances of some of them exceed 20 kpc.

In the catalog by Stanghellini and Haywood (2010) the distances to
PNe were estimated using a statistical method (Cahn et al. 1992;
Stanghellini et al. 2008). Data on the radio flux from
observations at a frequency of 5 GHz are used to calculate the
mass of the ionized shell. It is important to note two facts.
First, the catalog by Stanghellini and Haywood (2010) contains a
fairly homogeneous material in which only a few bipolar nebulae
poorly follow the PN mass--radius relations. Second, the distance
scale was calibrated on the basis of PNe from the Magellanic
Clouds. According to the estimate by Stanghellini et al. (2008),
the distances are determined in this way with a relative error of
$\sim$30\%. The actual error of these distance estimates can be
larger. As a comparative analysis by Smith (2015) showed, the
individual distance estimates for PNe could differ by a factor of
2--2.5.

%%%%%%%%%%%%%%%%%%%%%%%%%%%%%%%%%%%%%%%%%%%%%%%%%%%%%%%%%%%%%%%%% Fig 1
 \begin{figure} {\begin{center}
 \includegraphics[width=80mm]{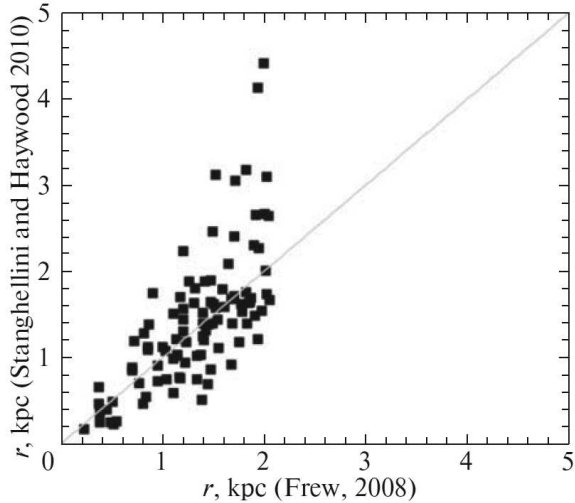}
 \caption{
Distances to PNe from the catalog by Stanghellini and Haywood
(2010) versus their distances from the catalog by Frew (2008).
 }
 \label{f-1} \end{center} } \end{figure}
%%%%%%%%%%%%%%%%%%%%%%%%%%%%%%%%%%%%%%%%%%%%%%%%%%%%%%%%%%%%%%%%%
%%%%%%%%%%%%%%%%%%%%%%%%%%%%%%%%%%%%%%%%%%%%%%%%%%%%%%%%%%%%%%%%% Fig 2
 \begin{figure} {\begin{center}
 \includegraphics[width=80mm]{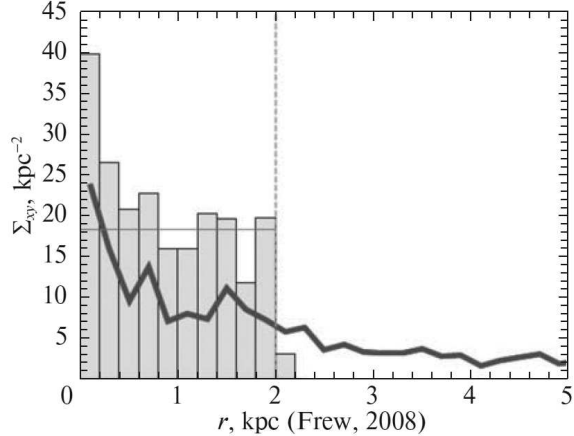}
 \caption{
Surface density distribution $\sum_{xy}$ of PNe from sample I as a
function of distance $d_{xy}$ in the form of a histogram, the
vertical dashed line indicates the adopted completeness boundary,
and the horizontal line indicates the mean surface density of this
sample; the surface density distribution of PNe from the catalog
by Stanghellini and Haywood (2010) is indicated by the thick solid
line.
 }
 \label{f-sum-2kpc} \end{center} } \end{figure}
%%%%%%%%%%%%%%%%%%%%%%%%%%%%%%%%%%%%%%%%%%%%%%%%%%%%%%%%%%%%%%%%%
%%%%%%%%%%%%%%             % Tab~1.
 \begin{table}[t]
 \caption[]
  {\small
 Parameters $z_\odot$ and $h_i,$~$i=1,2,3,$ found in this paper from the samples of
 PNe with Peimbert’s types I, II, and III
  }
  \begin{center}  \label{t:01}   %\small %\footnotesize\baselineskip=0.1ex
  \begin{tabular}{|c|c|c|c|l|c|}\hline
 $z_\odot,$~pc & $h_1,$~pc  & $h_2,$~pc  & $h_3,$~pc  & Sample                    & $N_\star$\\\hline
   $~-6\pm7~$  & $197\pm10$ & $168\pm7~$ & $196\pm9~$ & sample~I,~ $d_{xy}<2$~kpc & 230 \\

   $-22\pm14$  & $203\pm22$ & $163\pm14$ & $186\pm17$ & sample~II, $r<2.0$~kpc &  55 \\
   $-13\pm14$  & $271\pm20$ & $245\pm18$ & $291\pm22$ & sample~II, $r<3.0$~kpc &  94 \\
   $-37\pm12$  & $316\pm15$ & $286\pm15$ & $337\pm18$ & sample~II, $r<4.0$~kpc & 131 \\
   $-57\pm11$  & $339\pm14$ & $299\pm14$ & $349\pm17$ & sample~II, $r<5.0$~kpc & 160 \\
   $-84\pm11$  & $382\pm13$ & $325\pm12$ & $376\pm15$ & sample~II, $r<6.0$~kpc & 199 \\
 \hline
   $-23\pm14$  & $193\pm22$ & $155\pm14$ & $177\pm17$ & sample~II, $r<2.0$~kpc &  55 \\
   $-10\pm11$  & $221\pm16$ & $201\pm14$ & $239\pm18$ & sample~II, $r<3.0$~kpc &  94 \\
   $-40\pm9~$  & $256\pm12$ & $232\pm12$ & $275\pm15$ & sample~II, $r<4.0$~kpc & 131 \\
   $-45\pm9~$  & $283\pm14$ & $247\pm11$ & $287\pm13$ & sample~II, $r<5.0$~kpc & 160 \\
   $-67\pm9~$  & $305\pm14$ & $265\pm10$ & $306\pm12$ & sample~II, $r<6.0$~kpc & 199 \\
 \hline
 \end{tabular}\end{center}  % {\small Примечание.}
 \end{table}
%%%%%%%%%%%%%%%%%%%%%%%%%%%%%%%%%%%%%%%%%%%%%%%%%%%%%%%%%%%%%%%%%

To study the kinematic properties of PNe, we used the
line-of-sight velocities and proper motions gathered from various
published sources using the SIMBAD electronic database. Note that
the proper motions are available only for a small fraction of the
catalog, and they were taken mainly from such catalogs as
Hipparcos (1997), Tycho-2 (H\o g et al. 2000), and UCAC4
(Zacharias et al. 2013). For 16 PNe the proper motions were taken
from the first published Gaia Data Release (DR1) (Gaia
Collaboration, Brown et al. 2016).

The catalog by Frew (2008) with distance estimates for PNe is of
great interest. A comparative analysis by Smith (2015) showed its
high quality. It contains more than 200 PNe within $\sim$2 kpc of
the Sun. Initially, they produced a calibration sample of 120 PNe
whose distances were estimated by several methods; subsequently,
the remaining observations were tied to this sample. The distances
to the remaining nebulae were determined by a statistical method
based on an empirical $H\alpha$ surface brightness--radius
relation (Frew and Parker 2006; Frew et al. 2013). This relation
was calibrated on the basis of PNe from the Magellanic Clouds.
According to the estimate by Frew (2008), the distances determined
by this method have a relative error of $\sim$22\% for highly
excited shells.

For several PNe we used more reliable distance
estimates (with a relative error of less than 20%).
These include NGC 6853, NGC 7293, Abell 31, and DeHt 5 whose
trigonometric parallaxes were measured on the basis of Hubble
Space Telescope observations (Benedict et al. 2009). For the PNe K
3--35 (Tafoya et al. 2011) and IRAS 1828--095 (Imai et al. 2013)
we used the trigonometric parallaxes that were determined from
VLBI observations within the Japanese VERA Program. In addition,
we used the ground-based optical determinations of the
trigonometric parallaxes for ten PNe from Harris et al. (2007):
NGC 6720, Abell 74, HDW 4, SH 2--216, PuWe 1, Ton 320, Abell 21,
Abell 7, Abell 24, and PG 1034+001.

Note that the SIMBAD database provides evidence (with the
corresponding bibliographic references) that some of the objects
in the list by Stanghellini and Haywood (2010) are not PNe. For
example, G097.5+03.1, G093.5+01.4, G298.1--00.7, and G328.9--02.4
are HII regions, G125.9--47.0 is a star, G173.7--05.8 is a
reflection nebula, and G104.8--06.7 and G330.7+04.1 are emission
line stars. According to Frew et al. (2013), two stars from the
list by Harris et al. (2007), namely RE 1738+665 and PHL 932, are
not PNe. We excluded all these objects from consideration.

In Fig. 1 the distances to PNe in the scale of Stanghellini and
Haywood (2010) are plotted against the distances in the scale of
Frew (2008). The lopsided distribution of points at large
distances is explained by the fact that there are no nebulae with
distances of more than about 2.2 kpc in the catalog by Frew
(2008). It can be surmised that if distant objects were present in
the catalog by Frew (2008), then the distribution could be
symmetric relative to the diagonal plotted in the figure. Such a
symmetry is observed at least up to 2 kpc.

For the subsequent work we produced sample I with completeness. It
consists of the catalog by Frew (2008) to which we added the above
nebulae with measured trigonometric parallaxes and the nebulae
from the catalog by Stanghellini and Haywood (2010). A total of 29
nearby ($r<2$~kpc) nebulae were added to the catalog by Frew
(2008). Sample I contains a total of 230 objects no farther than 2
kpc from the Sun.

Figure 2 shows the surface density distribution $\sum_{xy}$ of PNe
as a function of distance $d_{xy}=r\cos b$. Based on this graph,
we assumed sample I to be complete up to distances of 2.0 kpc and
the mean surface density $\sum_{xy}$ to be 18.3 kpc$^{-2}.$

Figure 2 also shows the surface density distribution of PNe from
the catalog by Stanghellini and Haywood (2010) without any
additions. As can be seen from the figure, the catalog by
Stanghellini and Haywood (2010) loses to sample I with regard to
completeness. However, the catalog by Stanghellini and Haywood
(2010) has a big advantage: it has many stars at large distances.
For the above nebulae with measured trigonometric parallaxes we
took the distances calculated from the parallaxes, took into
account the misclassified nebulae, and called this sample II.
Sample II contains a total of 726 PNe.

To assign the PNe to particular Galactic subsystems, it is
convenient to use the classification introduced by Peimbert (1978)
and improved by Quireza et al. (2007). The nebulae of Peimbert's
types I, II, III, IV, and V belong to the thin disk, the thick
disk, the halo, and the bulge, respectively. For example,
according to Table 5 from Milanova and Kholtygin (2009), the scale
height for PNe of different types changes from $\sim$0.2 kpc for
type I to $\sim$1 kpc for type III and $\sim$1--2 kpc for type IV.
We took the values of the types from Quireza et al. (2007).
However, it turned out that information about the PN types
according to Peimbert's classification is available only for 60\%
of the PNe from the list by Stanghellini and Haywood (2010).

Note that due to the presence of PNe of different types in our
sample II, the scale heights $h_1, h_2,$ and $h_3$ can increase
with sample radius $r,$ because old halo objects fall into the
sample.

%%%%%%%%%%%%%%%%%%%%%%%%%%%%%%%%%%%%%%%%%%%%%%%%%%%%%%%%%%%%%%%%% Fig 3
 \begin{figure} {\begin{center}
 \includegraphics[width=150mm]{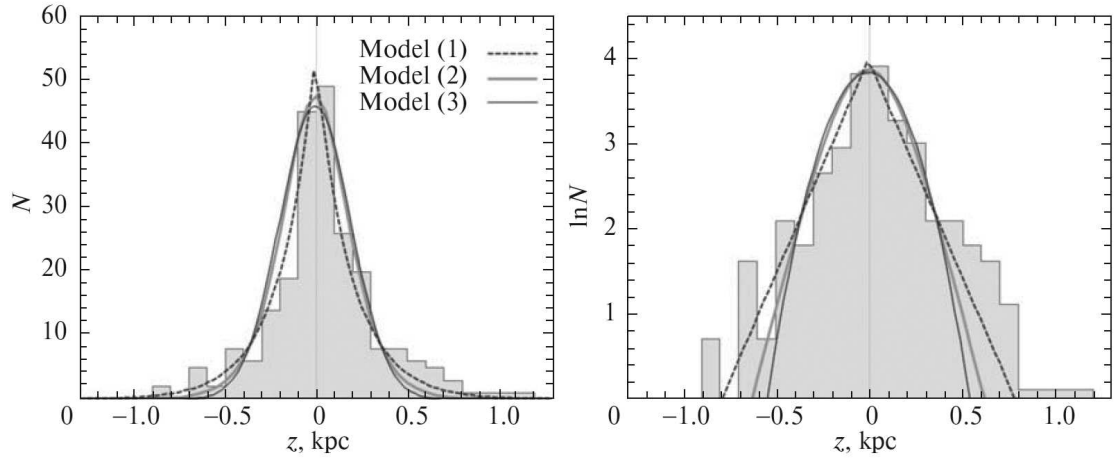}
 \caption{Histogram of the $z$ distribution for the PNe of sample I from the range of
 distances $d_{xy}<2$ kpc in linear (a) and logarithmic (b) scales.
 }
 \label{f-hist-2kpc} \end{center} } \end{figure}
%%%%%%%%%%%%%%%%%%%%%%%%%%%%%%%%%%%%%%%%%%%%%%%%%%%%%%%%%%%%%%%%%
%%%%%%%%%%%%%%%%%%%%%%%%%%%%%%%%%%%%%%%%%%%%%%%%%%%%%%%%%%%%%%%%% Fig 4
 \begin{figure} {\begin{center}
    \includegraphics[width=0.55\textwidth]{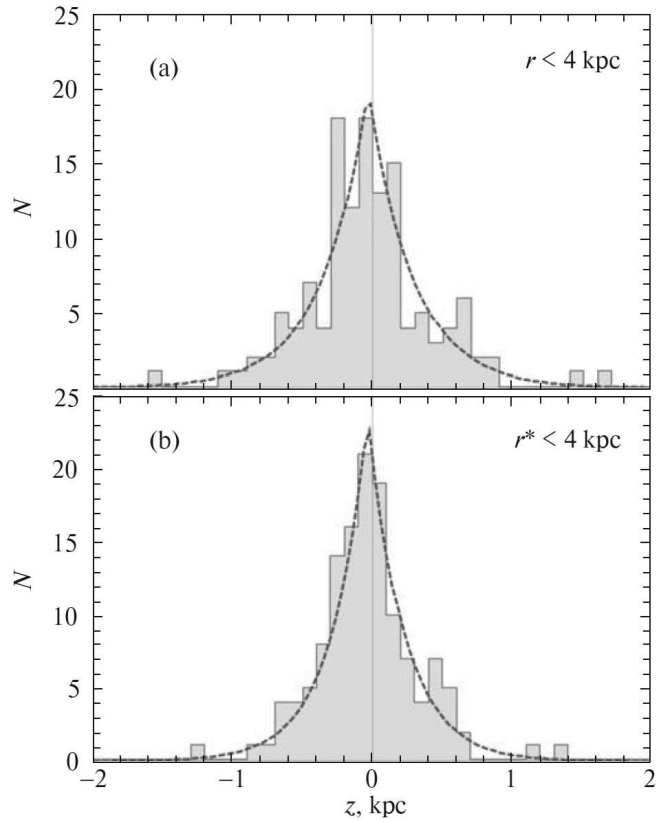}
 \caption{Histogramof the $z$ distributionfor the PNe of sample II with Peimbert's
 types I, II, and III (a) from the range of distances $r<4$~kpc; (b) with distances
 $r^*=0.8r$ reduced by 20\%.
 }
 \label{f-hist-II} \end{center} } \end{figure}
%%%%%%%%%%%%%%%%%%%%%%%%%%%%%%%%%%%%%%%%%%%%%%%%%%%%%%%%%%%%%%%%%
%%%%%%%%%%%%%%%%%%%%%%%%%%%%%%%%%%%%%%%%%%%%%%%%%%%%%%%%%%%%%%%%% Fig 5
 \begin{figure} {\begin{center}
    \includegraphics[width=0.55\textwidth]{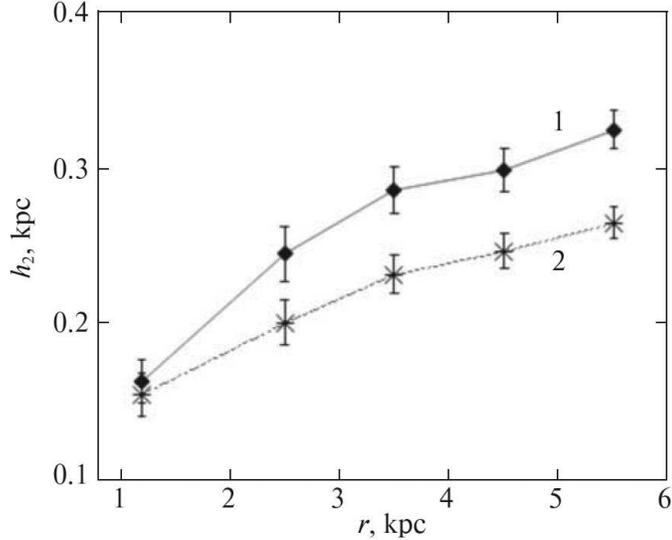}
 \caption{Scale height $h_2$ found from the PNe of sample II
with Peimbert's types I, II, and III (line 1) and from the same
nebulae with distances reduced by 20\% (line 2) versus
heliocentric distance $r.$
 }
 \label{f-5} \end{center} } \end{figure}
%%%%%%%%%%%%%%%%%%%%%%%%%%%%%%%%%%%%%%%%%%%%%%%%%%%%%%%%%%%%%%%%%

 \subsection*{RESULTS AND DISCUSSION}
Using our complete sample I with parameters $d_{xy}=2.0$~kpc,
$N_f=230,$ and $\sum_{xy}=18.3$ kpc$^{-2}$ based on Eq. (4), we
estimate the total number of PNe in the Galaxy to be
$N_{tot}=14929\pm2850.$

Table 1 gives the parameters $z_\odot$ and $h_i (i=1,2,3)$ found
from samples I and II. For sample I we used the additional
constraint $|z|<1.8$~kpc. Several solutions obtained with various
constraints on the radius $r$ are given in the case of analyzing
sample II. For sample II we used the additional constraint
$|z|<2.0$~kpc. Here, we use only those PNe from sample II that
have Peimbert’s types I, IIa, IIb, and III. In our sample there
are very few type III nebulae that, according to this
classification, belong to the thick disk, no more than ten.

The errors of the sought--for parameters were estimated through
statistical Monte Carlo simulations. The error estimates were made
by generating 1000 random realizations. The measurement errors
were added to the distances; we assumed the random errors of the
distances in the catalog by Frew (2008) to be 30\%.

The histogram of the $z$ distribution constructed using sample I
is shown in Fig. 3 in both linear and logarithmic scales. Figure 3
displays three curves constructed on the basis of models (1)--(3).
As can be seen from the figure, models (2) and (3) describe the
broad wings of the distribution more poorly than does model (1).

Note that Zijlstra and Pottasch (1991) used three density
distributions to determine the scale height based on PNe. These
authors found relations between the scale heights slightly
different from ours: $h_1 = 250$ pc, $h_2=380/\sqrt{2}=269$~pc
(given their modification of the law (2)), and $h_3=490$~pc. We
agree with the conclusions by Zijlstra and Pottasch (1991) that
(1) is the best law to analyze the vertical distribution of PNe.

There is a considerable growth of h with sample radius for the PNe
from sample II that is difficult to explain. As can be seen from
Table 1, for the sample from the range $r<6.0$~kpc the scale
height reaches $h_1=382$~kpc. Such a value is already typical of
thick disk objects. There is no reason to assume that the
admixture of PNe from the thick disk and the halo increases
considerably with distance in this sample. One of the solutions of
this problem is to reduce the distance scale of the catalog by
Stanghellini and Haywood (2010). We experimentally found that it
would suffice to reduce the distance scale approximately by 20\%
to obtain acceptable results. The results of such an approach are
presented in the lower part of Table 2 and in Figs. 4 and 5. For
this purpose, the distance to each star was scaled, and the entire
histogram was constructed again, the distances to the nebulae with
measured trigonometric parallaxes did not change.

Figure 4 provides the histograms constructed from the data of
Table 2 for the PNe of sample II with Peimbert’s types I, II, and
III selected under the condition $r<4$~kpc. These histograms were
constructed using both the original distances and those multiplied
by a factor of 0.8. The curves according to an exponential law
were fitted. As can be seen from the figure, reducing the
distances leads to a significant improvement of the histogram
shape.

In Fig. 5 the scale height $h_2$ is plotted against the
heliocentric distance $r.$ The values of $h_2$ were taken from
Table 1. We chose precisely $h_2,$ because their values are
determined with smaller random errors than those of $h_1$ and
$h_3.$ It can be clearly seen from this figure that a relatively
small reduction of the distance scale from Stanghellini and
Haywood (2010) causes the distance dependence of $h$ to decrease
(the slope of the curve to decrease).

 \subsubsection*{The Number of PNe in the Galaxy}
Using the parameters $L_d=3.3$~kpc and $R^{lim}=15.0$~kpc close to
those that we use with the surface density
$\sum_{xy}=23.4$~kpc$^{-2}$ found, Frew (2008) determined
$N_{tot}=19800\pm4000.$ The value of $N_{tot}=14929\pm2850$ we
found (essentially based on his catalog with minor additions) is
consistent with its estimate by Frew (2008) within $1\sigma.$ The
existing difference is explained mainly by the difference in the
surface densities $\sum_{xy}$ determined by different methods.
However, according to Frew (2008), one might also expect a larger
number of PNe in the Galaxy, $N_{tot}=24000\pm4000.$ Such a value
was obtained by taking into account the fact that the surface
density of PNe increases in the Galactic bulge region.

Following the approach of Frew (2008), we increased our estimate
of the total number of PNe by 2000, which takes into account the
density rise in the Galactic bulge. As a result, we obtained
$N_{tot}=16929\pm2850.$

%%%%%%%%%%%%%%       % Tab~2.
 \begin{table}[t]
 \caption[]  %\label{t2-Liter}
   {\small
 Parameters $z_\odot$ and $h_1$ found by various objects from thin-disk objects
  }
  \begin{center}  \label{t:02}   %\small %\footnotesize\baselineskip=0.1ex
  \begin{tabular}{|c|c|c|l|c|c|}\hline
 $z_\odot,$~pc & $h_1,$~pc & Author & Sample   \\\hline

   $-24\pm2$ & $~84\pm2~$  &  (1) & 250 Cepheids, $\sim$138~Myr, $r<4$~kpc \\
   $-15\pm2$ & $150\pm27$  &  (2) & open clusters, 200--1000 Myr \\
   $      0$ & $250\pm50$  &  (3) & PNe, $r<1$~kpc \\
   $      0$ & 259~~~~~    &  (4) & 196 PNe \\
   $      0$ & $217\pm20$  &  (5) & PNe, $r<2$~kpc \\
   $      0$ & $236\pm10$  &  (6) & AGB stars, $r<1.4$~kpc \\
  $-27~~~~~$ & 220--300    &  (7) & 717 white dwarfs \\
   $-27\pm4$ & $330\pm3~$  &  (8) & old thin-disk stars, SDSS \\
   $      0$ & 300~~~~~    &  (9) & $M$ dwarfs from SDSS catalog, $r<2$~kpc \\
 \hline
 \end{tabular}\end{center}  {\small
(1) Bobylev and Bajkova (2016a), (2) Bonatto et al. (2006), (3)
Zijlstra and Pottasch (1991), (4) Corradi and Schwarz (1995), (5)
Frew (2008), (6) Olivier et al. (2001),(7) Vennes et al. (2002),
(8) Chen et al. (2001), (9) Juri\'c et al. (2008).
   }
 \end{table}
%%%%%%%%%%%%%%%%%%%%%%%%%%%%%%%%%%%%%%%%%%%%%%%%%%%%%%%%%%%%%%%%%

 \subsubsection*{The Scale Height}
Table 2 gives a brief overview of $z_\odot$ and $h_1$ found by
various authors from Galactic thin-disk objects. The thin disk is
not a homogeneous formation. The first two rows of Table~2 give
the estimates obtained from intermediate-age objects. For example,
the scale height $h_1$ determined from the youngest fraction of
the thin disk ranges from $26.5\pm0.7$~pc for a sample of methanol
masers (Bobylev and Bajkova 2016b) to $50\pm3$~pc (HII regions,
Wolf--Rayet stars; Bobylev and Bajkova 2016a).

Bobylev and Bajkova (2016a) found $z_\odot$ and $h_1$ by analyzing
a sample of classical Cepheids with a mean age of 138 Myr. Open
star clusters with various ages served as the subject of analysis
in Bonatto et al (2006). For the oldest clusters these authors
failed to determine the vertical disk scale height. However,
Froebrich (2014) showed that for open clusters with an age of more
than 1~Gyr the scale height increases sharply ($h_1>200$~pc) and
rapidly reaches values that are more typical of thick-disk
objects, for example, $h_1=550$~pc for clusters with an age of 3.5
Gyr.

The vertical disk scale height has been determined repeatedly by
various authors from PNe. Zijlstra and Pottasch (1991) analyzed
the vertical distribution of 37 PNe within 1 kpc of the Sun.
Corradi and Schwarz (1995) considered a sample of 196 nearby PNe
with various morphologies. For example, these authors found the
smallest scale height $h_1=130$~pc from 35 bipolar nebulae and the
largest one $h_1=325$~pc from 119 elliptical nebulae. Note that
Frew (2008) determined h1 using data on PNe from his catalog with
the additional constraint $|z|\leq300$~pc. Interestingly, we
obtained a similar result (sample I) without such a strong
constraint.

Olivier et al. (2001) found $h_1=236\pm10$~pc from a sample of 58
AGB stars with medium and high mass loss rates. These stars have
masses in the range 1-–2$M_\odot$ and are the direct progenitors
of PNe. To estimate the distances to these stars, we used their
photometric characteristics in the infrared $JHKL$ bands and the
data on their variability.

Based on stars counts for two large samples of stars from the SDSS
(Sloan Digital Sky Survey, York et al. 2000) catalog, Chen et al.
(2001) constructed models for the thin and thick disks as well as
the Galactic halo. Juri\'c et al. (2008) confirmed the result of
Chen et al. (2001) by analyzing a large high-latitude
($|b|>25^\circ$) sample of M dwarfs from the SDSS catalog. This
catalog is distinguished by a high accuracy of determining the
photometric distances of stars. The mean error of the photometric
characteristics in it is $\sim$0.02$^m;$ the error in the absolute
magnitude $\sigma_{M_r}$ is $\sim$0.3$^m.$ Then, according to the
estimate by Juri\'c et al. (2008), the error in the photometric
distances of stars is $\sim$18\%.

Vennes et al. (2002) studied a sample of 942 spectroscopically
identified hydrogen-rich (DA) white dwarfs. They adopted
$z_\odot=-27$~pc according to the estimate by Chen et al. (2001).

The values of $h_1$ presented in Table 2 served us as a guide for
choosing the scale factor 0.8 of the distance scale of the catalog
by Stanghellini and Haywood (2010). Note that the various distance
scales of PNe have been compared by various authors repeatedly
(Ortiz 2013; Smith 2015), and a discrepancy between the scales of
20\% is encountered quite often.

The vertical scale height $h_1=197\pm10$~pc that we found based on
230 PNe from the combined sample (sample I) is in good agreement
with the results of other authors. Note that among the results of
the analysis of PNe presented in Table 2, our estimate was
obtained from the largest number of nebulae and is distinguished
by a high accuracy.

 \subsubsection*{The Galactic Rotation Parameters}
We know the various kinematic methods that allow the distance
scale factor of the sample being studied to be estimated. A
detailed description of several such methods can be found in
Bobylev and Bajkova (2011). For example, the approach where the
distance scale factor is included as an additional unknown in the
original kinematic equations (5)--(7) is efficient (Dambis et al.
2001; Zabolotskikh et al. 2002). In this case, it can be
determined by simultaneously solving the system of equations or
found by minimizing the residuals according to the $\chi^2$ test.
This method is based on the assumption that the velocities
directed along and perpendicularly to the line of sight are, on
average, equal. Unfortunately, sample II contains very few PNe
with measured proper motions; therefore, such and similar
approaches are difficult to implement at present. Another approach
based on an adjustment of the second derivative of Galactic
rotation $\Omega^{'}_0$ in external convergence is acceptable.

%%%%%%%%%%%%%%%%%%%%%%%%%%%%%%%%%%%%%%%%%%%%%%%%%%%%%%%%%%%%%%%%% Fig 6
 \begin{figure}[t] {\begin{center}
   \includegraphics[width=0.60\textwidth]{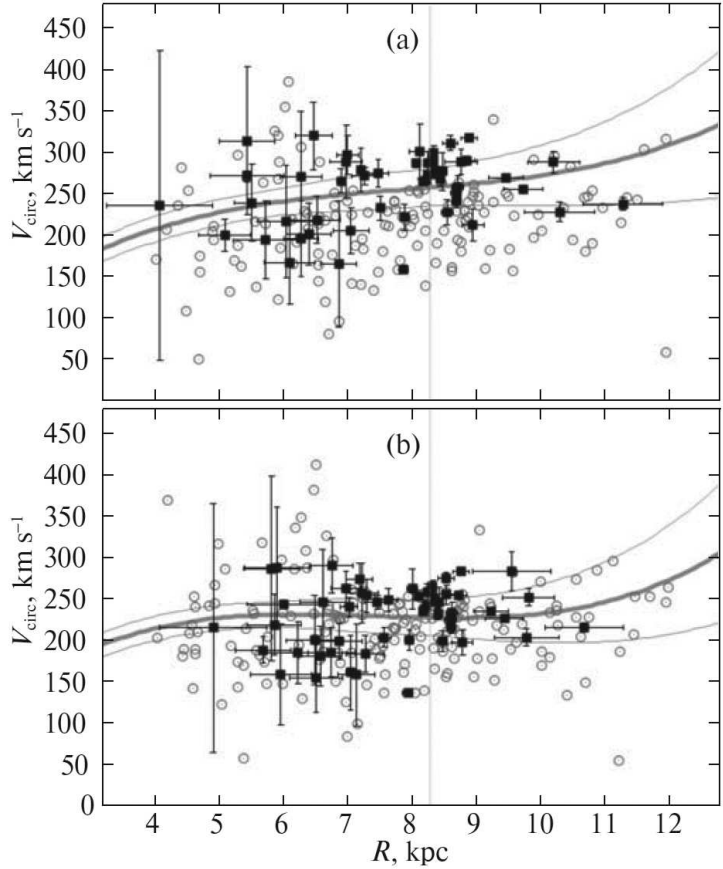}
 \caption{
(a) The Galactic rotation curve constructed from solution (9) with
an indication of the boundaries of the 1у confidence intervals,
(b) the Galactic rotation curve constructed from solution (10) in
which the distances to PNe reduced by 20\% were used; the vertical
dashed line marks the Sun's position, the filled squares with
error bars give the velocities of the PNe with measured
line-of-sight velocities and proper motions, the circles give the
velocities of the nebulae for which only their line-of-sight
velocities were measured.}
 \label{f-6} \end{center} } \end{figure}
%%%%%%%%%%%%%%%%%%%%%%%%%%%%%%%%%%%%%%%%%%%%%%%%%%%%%%%%%%%%%%%%%

To estimate the Galactic rotation parameters, we use the following
method. PNe with the proper motions, line-of-sight velocities, and
distances give all three Eqs. (5)--(7), while PNe only with the
line-of-sight velocities give only one Eq. (5).

No PNe of Peimbert's types IV and V were used. We limited our
sample by the radius $r=6$~kpc, $4<R<12$~kpc, and $|z|=2$~kpc. 170
nebulae only with the line-of-sight velocities and 56 nebulae with
complete information are involved in the solution; the total
number of equations is 382.

The system of conditional equations (5)--(7) is solved by the
least-squares method with weights
 $w_r=S_0/\sqrt {S_0^2+\sigma^2_{V_r}},$
 $w_l=S_0/\sqrt {S_0^2+\sigma^2_{V_l}}$ and
 $w_b=S_0/\sqrt {S_0^2+\sigma^2_{V_b}},$ where $S_0$ is the
``cosmic'' dispersion, $\sigma_{V_r}, \sigma_{V_l}, \sigma_{V_b}$
are the dispersions of the errors in the corresponding observed
velocities. The value of $S_0$ is comparable to the
root-mean-square residual у0 (the error per unit weight) when
solving the conditional equations (5)--(7); we adopted $S_0=45$ km
s$^{-1}$.

Using the original distances to
 \begin{equation}
 \label{solution-1}
 \begin{array}{lll}
 (U,V,W)_\odot=(16,26,9)\pm(3,5,5)~\hbox{km s$^{-1}$},\\
      \Omega_0=~31.1\pm2.7~\hbox{km s$^{-1}$ kpc$^{-1}$},\\
  \Omega^{'}_0=-2.98\pm0.32~\hbox{km s$^{-1}$ kpc$^{-2}$},\\
 \Omega^{''}_0=~0.84\pm0.28~\hbox{km s$^{-1}$ kpc$^{-3}$}.
 \end{array}
 \end{equation}
In this solution the error per unit weight is $\sigma_0=40.7$~km
s$^{-1}$, the Oort constants are $A=-12.4\pm1.3$ km s$^{-1}$
kpc$^{-1}$  and $B=18.8\pm3.0$ km s$^{-1}$ kpc$^{-1}$. The
solution was found after the elimination of very large residuals
according to the 3у criterion. The large random error in Щ0 is
explained by the fact that the value of this quantity can be
determined only from Eq. (6), i.e., only when using the proper
motions, but they are few and their errors are large, with these
errors (when converted to km s$^{-1}$) increasing with distance.

Comparison of the values of $\Omega_0$ found from Galactic masers
with measured trigonometric parallaxes,
 $\Omega^{'}_0=-3.96\pm0.09$ km s$^{-1}$ kpc$^{-2}$ at $R_0=8.3$~kpc (Bobylev and Bajkova 2014) and
 $\Omega^{'}_0=-3.96\pm0.07$ km s$^{-1}$ kpc$^{-2}$ at $R_0=8.4$~kpc (Rastorguev et al. 2016),
with the result of solution (9) gives the distance scale factor
2.98/3.47=0.75.

Once all distance were multiplied by the factor 0.8, we found the
following kinematic parameters:
 \begin{equation}
 \label{solution-2}
 \begin{array}{lll}
 (U,V,W)_\odot=(12,27,7)\pm(3,4,6)~\hbox{km s$^{-1}$},\\
      \Omega_0=~27.4\pm2.6~\hbox{km s$^{-1}$ kpc$^{-1}$},\\
  \Omega^{'}_0=-3.47\pm0.30~\hbox{km s$^{-1}$ kpc$^{-2}$},\\
 \Omega^{''}_0=~1.20\pm0.26~\hbox{km s$^{-1}$ kpc$^{-3}$}.
 \end{array}
 \end{equation}
In this solution the error per unit weight decreased in comparison
with the previous solution and is $\sigma_0=40.7$ km s$^{-1}$, the
Oort constants are $A=-14.4\pm1.2$ km s$^{-1}$ kpc$^{-1}$ and
$B=13.0\pm2.9$ km s$^{-1}$ kpc$^{-1}$. The linear rotation
velocity of the Galaxy near the Sun in solution (10) is
$V_0=\Omega_0 R_0=227\pm30$ km s$^{-1}$. Such a value of the
velocity $V_0$ is typical of the young fraction of the Galactic
disk. For example, by analyzing masers with measured trigonometric
parallaxes, Reid et al. (2014) found $V_0=240\pm8$~km s$^{-1}$
($R_0=8.34\pm0.16$~kpc), while Honma et al. (2012) obtained an
estimate of $V_0=238\pm14$ km s$^{-1}$ ($R_0=8.05\pm0.45$~kpc)
from a smaller number of masers. As a check, we obtain the
distance scale factor 2.98/3.47=0.86 by comparing the values of
$\Omega^{'}_0$ found in solutions (9) and (10).

Figure 6 presents two Galactic rotation curves constructed from
solutions (9) and (10). When there are only the line-of-sight
velocities for PNe, the components of their circular rotation
velocities $V_{circ}$ are calculated from the well-known formula
 \begin{equation}
  V_{circ}=|R\omega_0|+RV_r/(R_0\sin l\cos b),
 \end{equation}
It can be seen from this formula that at small values of $\sin l$
in the denominator these velocities have very large errors.
Therefore, when constructing Fig. 6, we excluded the PNe from the
cone $|l|<15^\circ.$ For PNe with known line-of-sight velocities
and proper motions we can calculate their total spatial
heliocentric velocities $U,V,$ and $W$ directed along the $x,y,$
and $z$ coordinate axes. For such PNe the rotation velocity
$V_{circ}$ is calculated from the relation
 \begin{equation}
  V_{circ}= U\sin \theta+(V_0+V)\cos \theta,
 \end{equation}
where $V_0=|R_0\omega_0|$ and the position angle $\theta$ is
defined as  $\tan\theta=y/(R_0-x)$.

As can be seen from Fig. 6, the Galactic rotation curve based on
the parameters of solution (10) is considerably closer to what is
obtained from the analysis of samples with more reliable distances
(Zabolotskikh et al. 2002; Bobylev and Bajkova 2011; Reid et al.
2014; Bobylev and Bajkova 2014; Rastorguev et al. 2016). Note that
the shape of the rotation curve in the solar neighborhood depends
on the Oort constants $A$ and $B,$ because $A+B=\partial
V_{circ}/\partial R$ at $R=R_0.$ In the case of solution (10),
 $A+B=-1.4$ km s$^{-1}$ kpc$^{-1}$ indicates that the Galactic rotation
velocity falls near the Sun, which is consistent with the results
of the analysis of other data. In contrast, in the case of
solution (9), the reverse is true: a large positive value of
 $A+B$ suggests a dramatic rise of the rotation curve, which is in poor
agreement with other data.

Thus, a kinematic analysis of sample II confirmed our previous
conclusion about the necessity of reducing the distance scale from
Stanghellini and Haywood (2010) approximately by 20\%.

The value of $\sigma_0=45$ km s$^{-1}$ (solution (10)) is the
velocity dispersion averaged over all directions. Using the
reduced distance scale, we calculated the dispersions of the
residual (the differential Galactic rotation and the peculiar
motion relative to the local standard of rest were taken into
account) velocities for the PNe from the entire sample II (the
nebulae of Peimbert’s types IV and V were excluded) as a function
of distance:
 $(\sigma_U,\sigma_V,\sigma_W)=(44,43,49)$ km s$^{-1}$ for the sample with $r<6$~kpc (40 nebulae),
 $(\sigma_U,\sigma_V,\sigma_W)=(44,37,32)$ km s$^{-1}$ for the sample with $r<3$ kpc (27 nebulae), or
 $(\sigma_U,\sigma_V,\sigma_W)=(44,35,33)$ km s$^{-1}$ for the sample with $r<2$~kpc (20 nebulae).
Note that the closer the sample to the Sun, the more reliable
values we obtain. It is interesting to compare these values, for
example, with the results of the analysis of 398 white dwarfs from
Pauli et al. (2006), where the following velocity dispersions were
found:
 $(\sigma_U,\sigma_V,\sigma_W)=(34,24,18)$ km s$^{-1}$ for a sample of thin-disk white dwarfs,
 $(\sigma_U,\sigma_V,\sigma_W)=(79,36,46)$ km s$^{-1}$ for a sample of thick-disk white dwarfs, and
 $(\sigma_U,\sigma_V,\sigma_W)=(138,95,47)$ km s$^{-1}$ for a sample of
halo white dwarfs. The velocity $V_\odot=27\pm4$ km s$^{-1}$ in
solution (10) suggests a slight lag behind the local standard of
rest due to an asymmetric drift, which provides evidence for the
relative youth of our sample of PNe.

Note that PNe are a set of ``stellar remnants'' with different
absolute ages. Therefore, the distribution of their spatial and
kinematic characteristics must correspond by 100\% neither to the
distribution of young open clusters nor to the properties of older
objects. We used the criteria that allowed the bulge and halo
objects to be excluded from our samples. Our analysis showed the
importance of using Peimbert's classification; therefore, in our
next publications we hope to use more fully the information about
the membership of PNe in particular Galactic subsystems.

 \subsection*{CONCLUSIONS}
Based on published data, we produced a sample of PNe that is
complete within 2 kpc of the Sun (sample I). The catalog by Frew
(2008), to which we added 29 PNe, served as a basis for this
sample. The additions include PNe with measured trigonometric
parallaxes and those from the catalog by Stanghellini and Haywood
(2010). We estimated the total number of PNe in the Galaxy from
this sample to be $N_{tot}=17 000\pm3000$ and determined the
vertical disk scale height based on an exponential density
distribution to be $h_1=197\pm10$~pc.

The second sample (sample II) includes PNe from the catalog by
Stanghellini and Haywood (2010) with minor additions. Based on
three density distributions, we found fairly large values of the
vertical scale height h from this sample. Here, we used only the
PNe with Peimbert's types I, IIa, IIb, and, in rare cases, III.
Thus, we took Galactic thin-disk objects in the overwhelming
majority of cases. For example, we found $h_1=316\pm15$~pc for PNe
from the range of distances $r<4$ kpc based on an exponential
density distribution. Here, we faced the fact that the values of h
increase considerably with sample radius. We experimentally found
that it is necessary to reduce the distance scale of the catalog
by Stanghellini and Haywood (2010) approximately by 20\% to obtain
acceptable results. In that case, for PNe from the range of
distances $r<4$~kpc the vertical scale height is
$h_1=256\pm12$~pc, while its values at greater distances are
consistent with the results of the analysis of other old Galactic
thin-disk objects, more specifically, it does not exceed 300 pc.

We provided sample II with published data on the line-of-sight
velocities and proper motions for the central stars of PNe. Based
on 226 PNe with reduced distances, we determined the following
Galactic rotation parameters:
 $(U,V,W)_\odot=(12,27,7)\pm(3,4,6)$ km s$^{-1}$,
 $\Omega_0=~27.4\pm2.6$ km s$^{-1}$ kpc$^{-1}$,
 $\Omega^{'}_0=-3.47\pm0.30$ km s$^{-1}$ kpc$^{-2}$, and
 $\Omega^{''}_0=~1.20\pm0.26$ km s$^{-1}$kpc$^{-3}$. The Oort constants
 $A=-14.4\pm1.2$ km s$^{-1}$ kpc$^{-1}$ and
 $B= 13.0\pm2.9$ km s$^{-1}$ kpc$^{-1}$ correspond to
this solution. The linear rotation velocity of the Galaxy at the
solar distance is $V_0=227\pm30$ km s$^{-1}$ for the adopted
$R_0=8.3$~kpc. The derived kinematic parameters are in good
agreement with those known from the literature. Our analysis of
$\Omega^{'}_0$ found confirmed the necessity of reducing the
original distance scale of the catalog by Stanghellini and Haywood
(2010) by 15--20\%.

 \medskip
We are grateful to the referees for their helpful remarks that
contributed to an improvement of this paper. This work was
supported by the Basic Research Program P--7 of the Presidium of
the Russian Academy of Sciences, the ``Transitional and Explosive
Processes in Astrophysics'' Subprogram.

 \bigskip\medskip{REFERENCES}\medskip{\small

1. A. Acker, F. Ochsenbein, B. Stenholm, R. Tylenda, J. Marcout,
and C. Schohn, in {\it Astronomy from Large Databases II,
Proceedings of the 43rd ESO Conference Workshop, Haguenau,
September 14--16, 1992}, Ed. by A. Heck and F. Murtagh (ESO,
1992), p. 163.

 2. D. Alloin, C. Cruz-Gonz\'alez, and M. Peimbert, Astrophys. J. 205, 74 (1976).

3. P.R. Amnuel, O.Kh. Guseinov, Kh.I. Novruzova, and Iu.S.
Rustamov, Astrophys. Space Sci. 107, 19 (1984).

 4. J.N. Bahcall and R.M. Soneira, Astrophys. J. Suppl. Ser. 44, 73 (1980).

 5. A.T. Bajkova and V.V. Bobylev, Baltic Astron. 24, 43 (2015).

6. G.F. Benedict, B.E. McArthur, R. Napiwotzki, T.E. Harrison,
H.C. Harris, E. Nelan, H.E. Bond, R.J. Patterson, and R.
Ciardullo, Astron. J. 138, 1969 (2009).

7. V.V. Bobylev and A.T. Bajkova, Astron. Lett. 37, 526 (2011).

8. V.V. Bobylev and A.T. Bajkova, Astron. Lett. 40, 389 (2014).

9. V.V. Bobylev and A.T. Bajkova, Astron. Lett. 42, 1 (2016a).

10. V.V. Bobylev and A.T. Bajkova, Astron. Lett. 42, 182 (2016b).

11. C. Bonatto, L.O. Kerber, E. Bica, and B.X. Santiago, Astron.
Astrophys. 446, 121 (2006).

 12. J. Bovy, H.-W. Rix, E.F. Schlafly, D.L. Nidever, J.A. Holtzman, M. Shetrone, and
     T.C. Beers, Astrophys. J. 823, 30 (2016).

 13. A.G.A. Brown, A. Vallenari, T. Prusti, J. de Bruijne, F. Mignard, R. Drimmel, et al.
    (GAIA Collab.), Astron. Astrophys. {\bf 595}, A2 (2016).

 14. A.S.M. Buckner and D. Froebrich, Mon. Not. R. Astron. Soc. 444, 290 (2014).

15. J.H. Cahn, J.B. Kaler, and L. Stanghellini, Astron. Astrophys.
Suppl. Ser. 94, 399 (1992).

16. B. Chen, C. Stoughton, J.A. Smith, A. Uomoto, J.R. Pier, B.
Yanny, \'Z.E. Ivezi\'c, D.G. York, et al., Astrophys. J. 553, 184
(2001).

 17. P.S. Conti and W.D. Vacca, Astron. J. 100, 431 (1990).

 18. R.L.M. Corradi and H.E. Schwarz, Astron. Astrophys. 293, 871 (1995).

19. A.K. Dambis, A.M. Mel’nik, and A.S. Rastorguev, Astron. Lett.
27, 58 (2001).

20. D.J. Frew and Q.A. Parker, in {\it Planetary Nebulae in our
Galaxy and Beyond, Proceedings of the 234th IAU Symposium}, Ed. by
M.J. Barlow and R.H. M\'endez (Cambridge Univ. Press, Cambridge,
2006), p. 49.

21. D.J. Frew, PhD Thesis (Depart. Phys., Macquarie Univ., NSW
2109, Australia, 2008).

22. D.J. Frew, I.S. Boji\v{c}i\'c, and Q. A. Parker,Mon. Not. R.
Astron. Soc. 431, 2 (2013).

23. S. Gillessen, F. Eisenhauer, T.K. Fritz, H. Bartko, K.
Dodds-Eden, O. Pfuhl, T. Ott, and R. Genzel, Astroph. J. 707, L114
(2009).

24. L. Girardi, M.A.T. Groenewegen, E. Hatziminaoglou, and L. da
Costa, Astron. Astrophys. 436, 895 (2005).

25. H.C. Harris, C.C. Dahn, B. Canzian, H.H. Guetter, S.K.
Leggett, S.E. Levine, C.B. Luginbuh, A.K.B. Monet, et al., Astron.
J. 133, 631 (2007).

26. The Hipparcos and Tycho Catalogues, ESA SP--1200 (1997).

27. E. Hog, C. Fabricius, V.V. Makarov, U. Bastian, P.
Schwekendiek, A. Wicenec, S. Urban, T. Corbin, and G. Wycoff,
Astron. Astrophys. 355, L27 (2000).

28. M. Honma, T. Nagayama, K. Ando, T. Bushimata, Y.K. Choi, T.
Handa, T. Hirota, H. Imai, T. Jike, et al., Publ. Astron. Soc.
Jpn. 64, 136 (2012).

 29. H. Imai, T. Kurayama, M. Honma, and T. Miyaji, Publ. Astron. Soc. Jpn. 65, 28 (2013).

 30. K. Ishida and R. Weinberger, Astron. Astrophys. 178, 227 (1987).

31. G.H. Jacoby, Astrophys. J. Suppl. Ser. 42, 1 (1980).

32. M. Juri\'c, Z. Ivezi\'c, A. Brooks, R.H. Lupton, D. Schlegel,
D. Finkbeiner, N. Padmanabhan, N. Bond, et al., Astrophys. J. 673,
864 (2008).

 33. G.S. Khromov, Space Sci. Rev. 51, 339 (1989).

 34. Yu. V.Milanova and A.F. Kholtygin, Astron. Lett. 35, 518 (2009).

 35. M. Moe and O. de Marco, Astrophys. J. 650, 916 (2006).

 36. E.A. Olivier, P. Whitelock, and F. Marang, Mon. Not. R. Astron. Soc. 326, 490 (2001).

 37. R. Ortiz, Astron. Astrophys. 560, A85 (2013).

38. Q. A. Parker, A. Acker, D.J. Frew, and W.A. Reid, in {\it
Planetary Nebulae in our Galaxy and Beyond, Proceedings of the
234th IAU Symposium}, Ed. by M. J. Barlow and R.H. Mendez
(Cambridge Univ. Press, Cambridge, 2006), p. 1.

39. E.-M. Pauli, R. Napiwotzki, U. Heber, M. Altmann, and M.
Odenkirchen, Astron. Astrophys. 447, 173 (2006).

40. M. Peimbert, IAU Symp. 76, 215 (1978).

41. M. Peimbert, Rev. Mex. Astron. Astrofis. 20, 119 (1990).

42. J.P. Phillips, Mon. Not. R. Astron. Soc. 353, 589 (2004).

43. A.E. Piskunov, N.V. Kharchenko, S. R\"oser, E. Schilbach, and
R.-D. Scholz, Astron. Astrophys. 445, 545 (2006).

44. T. Prusti, J.H.J. de Bruijne, A.G.A. Brown, A. Vallenari, C.
Babusiaux, C.A.L. Bailer-Jones, U. Bastian, M. Biermann, et al.
(GAIA Collab.), Astron. Astrophys. {\bf 595}, A1 (2016).

45. C. Quireza, H.J. Rocha-Pinto, and W.J. Maciel, Astron.
Astrophys. 475, 217 (2007).

46. A.S. Rastorguev, M.V. Zabolotskikh, A.K. Dambis, N. D. Utkin,
A.T. Bajkova, and V.V. Bobylev, arXiv: 1603.09124 (2016).

47. M.J. Reid, K.M. Menten, A. Brunthaler, X.W. Zheng, T.M. Dame,
Y. Xu, Y. Wu, B. Zhang, et al., Astrophys. J. 783, 130 (2014).

48. H.-W. Rix and J. Bovy, Astron. Astrophys. Rev. 21, 61 (2013).

49. A.C. Robin, C. Reyl\'e, S. Derri\'ere, and S. Picaud, Astron.
Astrophys. 409, 523 (2003).

50. I.S. Shklovskii, Astron. Zh. 33, 222 (1956).

51. H. Smith, Mon. Not. R. Astron. Soc. 449, 2980 (2015).

52. L. Stanghellini, R.A. Shaw, and E. Villaver, Astrophys. J.
689, 194 (2008).

 53. L. Stanghellini and M. Haywood, Astrophys. J. 714, 1096 (2010).

 54. L. Stanghellini, B. Bucciarelli, M.G. Lattanzi, and R. Morbidelli,
     arXiv: 1609.08840 (2016).

55. D. Tafoya, H. Imai, Y. Gomez, J.M. Torrelles, N.A. Patel, G.
Anglada, L.F. Miranda, M. Honma, et al., Publ. Astron. Soc. Jpn.
63, 71 (2011).

56. S. Vennes, R.J. Smith, B.J. Boyle, S.M. Croom, A. Kawka, T.
Shanks, L. Miller, and N. Loaring, Mon. Not. R. Astron. Soc. 335,
673 (2002).

57. D.G. York, J. Adelman, J.E. Anderson, S.F. Anderson, J. Annis,
N.A. Bahcall, J.A. Bakken, R. Barkhouser, et al., Astrophys. J.
540, 825 (2000).

58. M.V. Zabolotskikh, A.S. Rastorguev, and A.K. Dambis, Astron.
Lett. 28, 454 (2002).

59. N. Zacharias, C.T. Finch, T.M. Girard, A. Henden, J.L.
Bartlett, D.G. Monet, and M.I. Zacharias, Astron. J. 145, 44
(2013).

60. A.A. Zijlstra and S.R. Pottasch, Astron. Astrophys. 243, 478
(1991).
 }

 \end{document}